\documentclass{PoS}
\usepackage{cite,amsmath}

\title{Inclusive Cross Sections in ME+PS Merging
\hfill
\begin{tiny}
DESY 13-118
\end{tiny}}

\ShortTitle{Inclusive Cross Sections in ME+PS Merging}

\author{\speaker{Simon Pl\"atzer}\\
        Deutsches Elektronen-Synchrotron (DESY),
        Notkestrasse 85, D-22607 Hamburg, Germany\\
        E-mail: \email{simon.plaetzer@desy.de}}

\abstract{We discuss an extension of matrix element plus parton shower
  merging at leading and next-to-leading order. The algorithm does
  preserve inclusive cross sections at the respective input
  order. This constraint avoids potentially large logarithmic
  contributions, which would require approximate (N)NLO contributions
  to cancel against.}

\FullConference{XXI International Workshop on Deep-Inelastic
  Scattering and Related Subject -DIS2013,\\ 22-26 April
  2013\\ Marseilles,France}

\begin{document}

\section{Introduction}

Parton shower event generators are by now indispensable workhorses of
experimental as well as theoretical studies for comparing (standard
model) theoretical predictions to measured observables in a most
detailed way.  Recent years have shown tremendous developments to
improve the accuracy of those simulations from the lowest order
description of hard scatterings to include exact tree level matrix
elements for a large number of jets as pioneered in
\cite{Catani:2001cc,Lonnblad:2001iq,Krauss:2002up}, and NLO
corrections for the hard scattering itself, \cite{Frixione:2002ik,
  Nason:2004rx}. Parton shower algorithms themselves have
undergone improvements in various directions, see {\it e.g.}
\cite{Schumann:2007mg,Giele:2007di,Platzer:2011bc,Kilian:2011ka,Platzer:2012qg}
and references therein, which have mainly been driven by easing the
combination of parton showers and higher order QCD corrections.
Following first steps of combining NLO corrections to the lowest order
process while including higher multiplicity tree level matrix elements
\cite{Hamilton:2010wh,Hoche:2010kg}, the combination of parton showers
and NLO corrections for several jets has been achieved
\cite{Lavesson:2008ah,Hoeche:2012yf,Frederix:2012ps}, partly by
including analytic resummation at high logarithmic order
\cite{Alioli:2012fc}.  In this contribution, we discuss algorithms to
combine parton showers and leading (LO) or next-to-leading order (NLO)
calculations for the production of multiple, additional jets,
focusing on the accuracy at which inclusive cross sections are
reproduced \cite{Platzer:2012bs,Lonnblad:2012ng,Lonnblad:2012ix}.

\section{The Landscape of Perturbative QCD}

Jet spectra, as well as event shapes and similar, infrared sensitive
observables at colliders exhibit a clear structure of leading
contributions at higher orders in the strong coupling. Particularly,
the perturbative expansion in terms of the strong coupling $\alpha_s$
is typically plagued by large logarithms of the infrared sensitive
quantity which eventually overcome the smallness of $\alpha_s$ and so
render fixed order predictions unreliable. The complications in these
regions confront with a certain simplicity of QCD amplitudes which
factor in the respective limits into universal and few process
dependent pieces at all orders, thus allowing the application of
resummation techniques to cure the convergence of the perturbative
series.

\begin{figure}
\begin{center}
\includegraphics[scale=0.3]{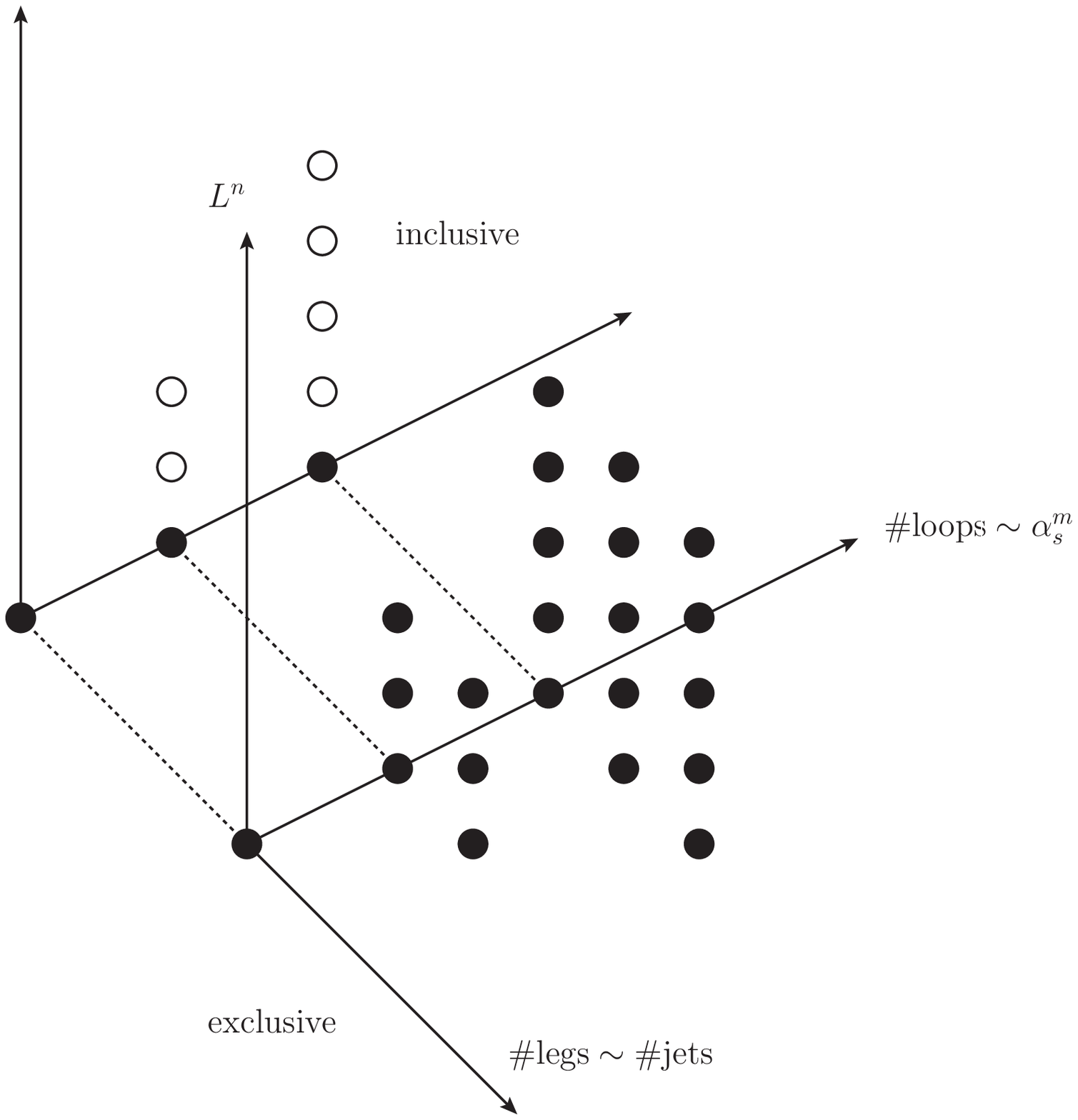}\hspace*{1cm}
\includegraphics[scale=0.3]{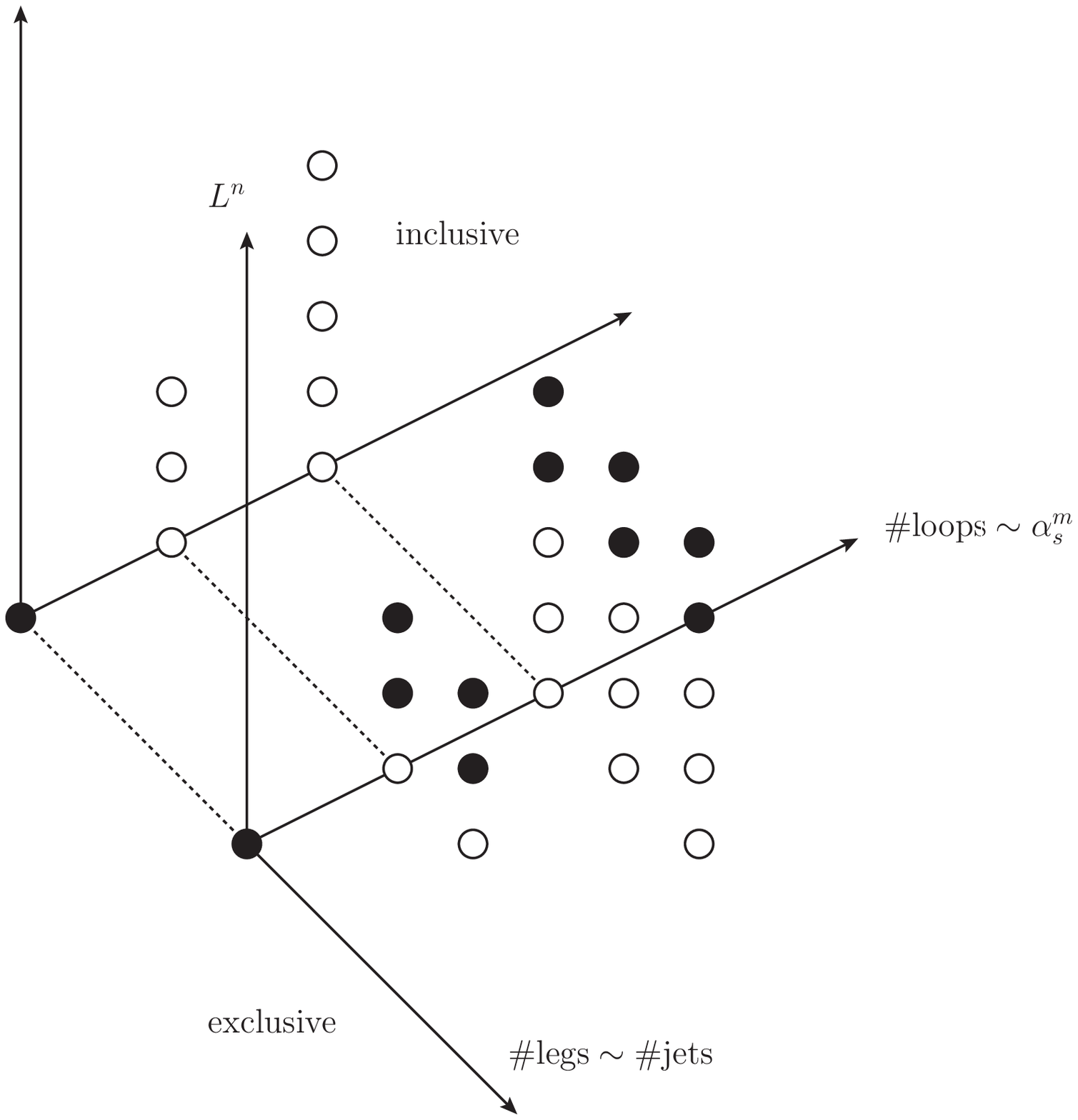}
\end{center}
\caption{\label{figures:landscape}Illustration of leading
  contributions to infrared sensitive observables in QCD perturbation
  theory. At each order in the strong coupling, large logarithmic
  contributions appear which cancel in inclusive quantities (left
  panel). The vertical axis indicates powers of large logarithms in
  dependence of the order in the strong coupling and the number of
  resolved or unresolved parton emissions (horizontal
  axes). Resummation at a certain logarithmic accuracy sums
  parts of these contributions to all orders (left panel, exemplifying
  a NLL resummation).}
\end{figure}

Notably in QCD (as opposed to electroweak corrections), inclusive
quantities are free of large logarithmic corrections with the regions
of resolved and unresolved parton emission, or exclusive and inclusive
jet cross sections to name an example, delivering equal in magnitude
and opposite in sign contributions of large logarithmic
enhancement. This situation is illustrated in
figure~\ref{figures:landscape}, along with an illustration of
resummation at next-to-leading logarithmic (NLL) accuracy, taking into
account all terms of the form $\alpha_s^n L^{2n}$ and $\alpha_s^n
L^{2n-1}$. Figure~\ref{figures:landscape} also indicates the origin of
the different contributions in terms of resolved (emission), and
unresolved (virtual), contributions which directly relate to the number
of legs and the order in the strong coupling considered.

\section{Parton Showers}

Parton showers simulate the multiple emission of soft and collinear
partons originating from a hard process configuration $\phi_n$ driven
by a differential cross section ${\rm
  d}\sigma(\phi_n,q_n|\cdots|q_0)$\footnote{See \cite{Platzer:2012bs}
  for details on the notation}. The sequence of scales $q_n < q_{n-1},
... , q_0$ is either directly determined from shower evolution, or is
assigned by a clustering procedure corresponding to the inverse of a
possible shower evolution $\phi_0\to \phi_n$, if the cross section is
determined by exact matrix elements. Sensible parton shower algorithms
can up to date only be proven to perform resummation at the level of
leading logarithms, but there are strong indications that a large part
of the next-to-leading logarithmic contributions are present, at least
in special cases. Parton showering acts on cross sections driving
events $\phi_n$ as
\begin{equation}
{\rm PS}_\mu\left[ {\rm d}\sigma(\phi_n,q_n)\right] =
{\rm d}\sigma(\phi_n,q_n) \Delta_n(\mu|q_n) +
{\rm PS}_\mu\left[{\rm d}\sigma(\phi_n,q_n)
\frac{{\rm d}\phi_{n+1}}{{\rm d}\phi_{n}}P_\mu(\phi_n,q_{n+1})\Delta_n(q_{n+1}|q_n)\right]\ ,
\end{equation}
where
\begin{equation}
\Delta_n(q|Q) = \exp\left(-\int_q^Q {\rm d}k\ \frac{{\rm d}\phi_{n+1}}{{\rm d}\phi_{n}{\rm d}k}\ P(\phi_n,k)\right) \ ,
\end{equation}
is the Sudakov form factor which can be interpreted as the probability
for no emission between a hard scale $Q$ and a soft scale $q$. The
cancellation of large logarithmic contributions in inclusive
quantities is inherent to parton shower evolution by the fact that
this evolution is a stochastic process being subject to unitarity
constraints,
\begin{equation}
\label{eqs:unitarity}
\int_q^{q_{k-1}}{\rm d}q_k  \frac{{\rm d}\phi_{k}}{{\rm d}\phi_{k-1}{\rm d}q_k}P(\phi_{k-1},q_{k})\Delta_{k-1}(q_k|q_{k-1}) =
1 - \Delta_{k-1}(q|q_{k-1}) \ .
\end{equation}
Considering inclusive and exclusive cross sections of emitting at
least $n$ or exactly $n$ resolved partons, parton shower evolution
will generate the cross sections
\begin{eqnarray}\nonumber
= n &\qquad& {\rm d}\sigma^{(0)}(\phi_0,q_0)
\frac{{\rm d}\phi_{n}}{{\rm d}\phi_{0}}P_\mu(\phi_{n-1},q_{n})\cdots
P_\mu(\phi_0,q_{1}) \Delta_{n}(\mu|q_{n}|\cdots|q_0)\\\nonumber
\ge n & \qquad & {\rm d}\sigma^{(0)}(\phi_0,q_0)
\frac{{\rm d}\phi_{n}}{{\rm d}\phi_{0}}P_\mu(\phi_{n-1},q_{n})\cdots
P_\mu(\phi_0,q_{1}) \Delta_{n-1}(q_{n}|\cdots|q_0) \ .
\end{eqnarray}
We will take this observation as a starting point to constrain
inclusive cross sections when combining parton showers and higher
multiplicity jet cross sections at leading order (LO) and
next-to-leading order (NLO).

\section{Tree Level Merging Revisited}

\begin{figure}
\begin{center}
\includegraphics[scale=0.3]{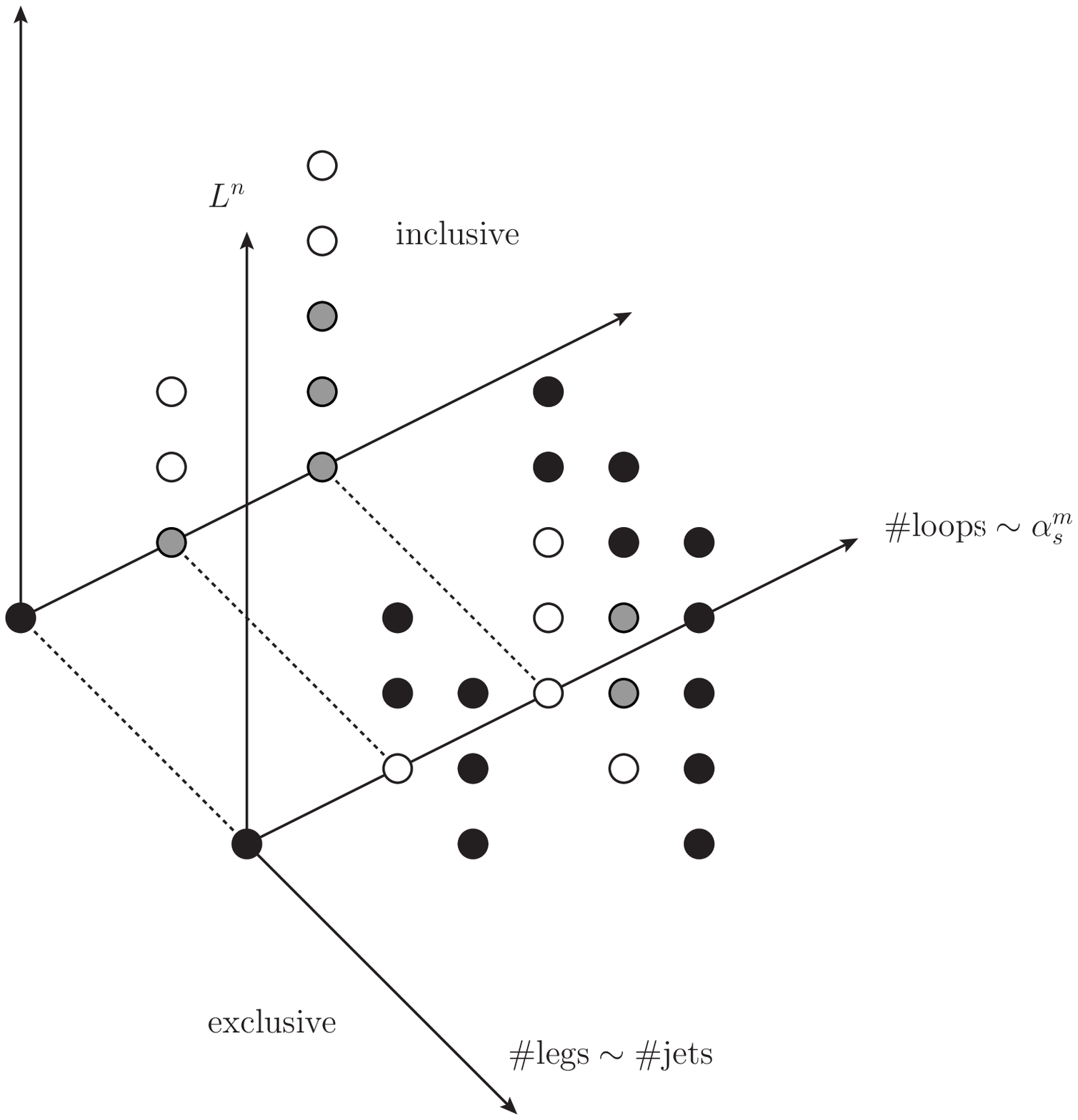}\hspace*{1cm}
\includegraphics[scale=0.3]{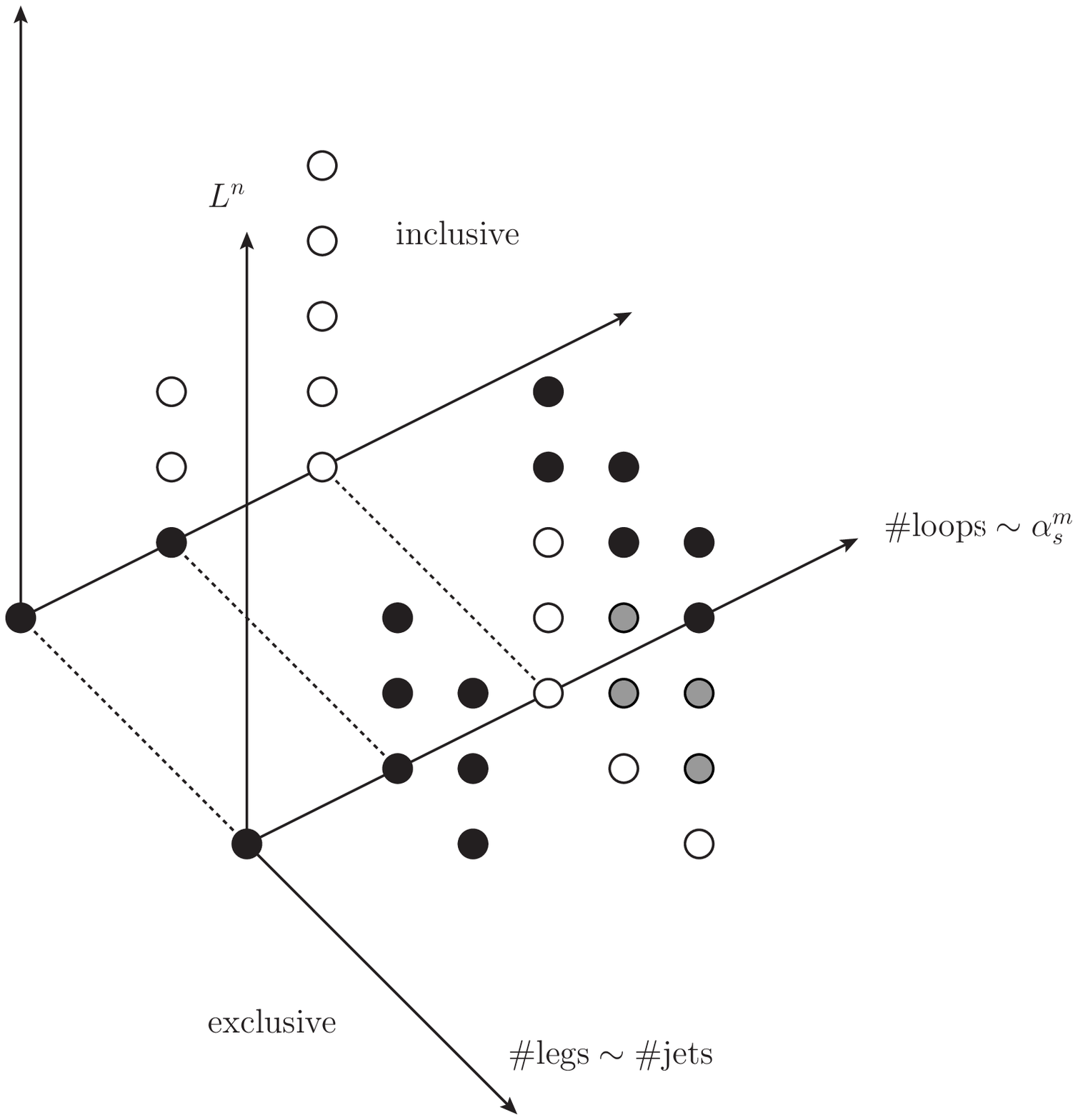}
\end{center}
\caption{\label{figures:mergingmatching}Merging tree level matrix
  elements and parton showers (left panel) allows to the exact
  description of subleading logarithmic contributions stemming from
  resolved, tree level contributions (up to two additional parton
  emissions in the case considered here). This will yield a partially
  improved description of contributions which would only be accessible
  from NLO calculations and may leave its traces in inclusive
  quantities (see text for details). This is to be confronted with NLO
  matching (right panel), which includes tree level contributions for
  one additional parton emission, as well as the exact one-loop pieces
  without modifying the inclusive cross section except adding the
  first order correction to it.}
\end{figure}

The merging of tree level matrix elements and parton showers aims at
improving the description of multiple parton emission by the
respective, exact, tree level matrix elements as far as
computationally feasible. This may either be done across all of phase
space or only in a certain region for large-angle, hard emissions.
Generically, we impose a merging condition to replace the product of
lowest order cross section and parton shower splitting probabilities
by their exact counterpart in exclusive cross sections generated by
the merging algorithm, while keeping the Sudakov form factor
accounting for exclusiveness of the respective final state.  Assuming
the presence of tree level matrix elements for emission of up to $N$
additional partons, and introducing a `hard' region as resolved by a
merging cutoff $\rho$ above the shower cutoff $\mu$,
we have
\begin{multline}
\label{eqs:mergedmerged}
{\rm PS}_\mu\left[{\rm d}\sigma^{\text{merged}}_{N,\rho}\right] =\\
{\rm PS}_{\mu | \rho}\left[
\sum_{k=0}^{N-1} {\rm d}\sigma_\rho^{(0)}(\phi_k,q_k)\Delta_{k}(\rho|q_k|\cdots |q_0)
+{\rm PS}_\rho\left[ {\rm d}\sigma_\rho^{(0)}(\phi_N,q_N)\Delta_{N-1}(q_{N}|\cdots |q_0) \right]\right] \ ,
\end{multline}
{\it i.e.} shower emissions are restricted to a range between the
merging scale $\rho$ and the infrared cutoff $\mu$ (as indicated by
the symbol ${\rm PS}_{\mu|\rho}$) for all but the highest matrix
element multiplicity available, and otherwise unconstrained. The
general solution to this condition is given in \cite{Platzer:2012bs},
and involves subtraction of the parton shower approximation in the
hard region above $\rho$, accompanied by an appropriate Sudakov
reweighting. This is illustrated in
figure~\ref{figures:mergingmatching} and compared to NLO matching.

\section{Inclusive Cross Sections as a Guide to NLO Merging}

By definition of the tree level merging condition, exclusive cross
sections resemble the functional form expected from parton shower
evolution; since no modification of the Sudakov form factors has been
made to restore the unitarity of the merging algorithm, inclusive
cross sections do, however, not reflect the expected pattern except
for the highest matrix element multiplicity available. Starting from
the next-to-maximal multiplicity, we instead find for the $n-1$ parton
inclusive cross section:
\begin{multline}\nonumber
 {\rm
  d}\sigma_\rho^{(0)}(\phi_{n-1},q_{n-1})\Delta_{n-2}(q_{n-1}|\cdots
|q_0) +\\\int_\rho^{q_{n-1}} {\rm d}q_n \left( \frac{{\rm
    d}\sigma_\rho^{(0)}(\phi_n,q_n)}{{\rm d}q_n}-\frac{{\rm
    d}\phi_{n}}{{\rm d}\phi_{n-1}{\rm d}q_n}
P_\rho(\phi_{n-1},q_n){\rm d}\sigma_\rho^{(0)}(\phi_{n-1},q_{n-1})
\right)\Delta_{n-1}(q_n|\cdots|q_0) \ .
\end{multline}
The first term is the expected result, while the correction to it
involves an integral over the difference between an $n$ parton cross
section and the approximation to it when replacing the softest
emission to be driven by the parton shower. Provided that the parton
shower is a reasonable approximation to the singly unresolved
behavior of tree level matrix elements, this contribution will not
contribute logarithmically enhanced contributions to inclusive cross
sections. The situation for merging parton showers and NLO corrections
of different jet multiplicity will be different. Here, the tree level
({\it i.e.} LO) exclusive $n$-parton cross sections (according to the
shower resolution with resolution parameter $\rho$) will be replaced
by their NLO counter parts as achieved {\it e.g.} by the algorithms
outlined in
\cite{Lavesson:2008ah,Hoeche:2012yf,Frederix:2012ps},
while the shower kernels still reflect the unresolved behavior of
tree level matrix elements, thus contributing higher logarithmic order
terms to inclusive quantities.

As a first step towards NLO merging with inclusive cross sections
which are not out of control, we therefore attempt to restore the LO
inclusive cross sections within a tree level merging algorithm, first.
Having identified the contribution by which inclusive cross sections
are spoiled, it is straightforward to subtract these contributions
from the merged cross section in an algorithmic way: Within the merged
cross section, for all but the highest multiplicity, the integral over
the phase space for the softest emission in the $n+1$ parton
contribution needs to be subtracted from the $n$ parton
contribution. This algorithm amounts to a variant of thew LoopSim
method \cite{Rubin:2010xp} and in the case of tree level merging
does generate approximate one-loop contributions, thus rendering the
LO merging effectively an nLO merging.

Expanding exclusive $n$-parton cross sections above the merging scale
$\rho$ to first order in the strong coupling, one readily identifies
the correction to promote the nLO cross section to a full NLO cross
section. The correction is given by the NLO calculation differential
in the Born variables, which will otherwise be treated in the same way
as the tree level input to the merging algorithm. The problematic
contributions in inclusive cross sections, as discussed previously,
are now manifest. We can, however, add correction terms similar to the
ones used for tree level merging to restore the exact NLO inclusive
cross section times the appropriate Sudakov suppression in a similar
algorithm. This procedure will generate approximate two loop
contribution, effectively rendering the algorithm a merging of LoopSim
nNLO calculations.

\section{Conclusions and Outlook}

We have presented an extension to multileg matrix element and parton
shower merging, which preserves inclusive cross sections at the level
of the available accuracy, particularly at tree and one-loop
level. This constraint is of utmost importance particularly for the
latter case, as NLO corrections to shower splitting kernels are so far
out of reach. The procedure starts from merging LO calculation of
different jet multiplicities, will generate approximate NLO
corrections similar to the LoopSim algorithm when imposing constraints
on inclusive cross sections and thus serves as a guide to NLO
merging. Repeating the procedure in this case yields approximate NNLO
corrections and we anticipate that this could well serve as a guide to
NNLO plus parton shower matching. An implementation of this algorithm
based on the Herwig++ Matchbox framework \cite{Platzer:2011bc} is
underway.

\section*{Acknowledgments}

This work was supported by the Helmholtz Alliance ``Physics at the
Terascale''.

\bibliography{multijets}

\end{document}